\documentclass[12pt,showpacs,twocolumn, preprintnumbers,aps,pre, reprint,superscriptaddress]{revtex4-1}

\usepackage{graphicx}
\usepackage{subfigure}
\usepackage{color}
\usepackage{float}
\usepackage{amsmath}
\usepackage[linktocpage,colorlinks=true,linkcolor=blue,citecolor=blue,breaklinks=true]{hyperref}
\usepackage{verbatim}
\usepackage{breakcites}

\newcommand{\RN}[1]{%
  \textup{\uppercase\expandafter{\romannumeral#1}}%
}

\begin{document}

\preprint{}

\title{Exoplanet Atmosphere Retrieval from Multifractal Analysis of Secondary Eclipse Spectra}

\author{Sahil Agarwal}
\affiliation{Yale University, New Haven, USA}

\author{John S. Wettlaufer}
\email[]{john.wettlaufer@su.se}
\affiliation{Yale University, New Haven, USA}
\affiliation{Nordita, Royal Institute of Technology and Stockholm University, SE-10691 Stockholm, Sweden}

\date{\today}

\begin{abstract}

We extend a data-based model-free multifractal method of exoplanet detection to probe exoplanetary atmospheres.  Whereas the transmission spectrum is studied during the primary eclipse, we analyze the \emph{emission spectrum} during the secondary eclipse, thereby probing the atmospheric limb. 
In addition to the spectral structure of exoplanet atmospheres, the approach provides information to study phenomena such as atmospheric flows, tidal-locking behavior, and the dayside-nightside redistribution of energy.  The approach is demonstrated using Spitzer data for exoplanet HD189733b. The central advantage of the method is the lack of model assumptions in the detection and observational schemes.

\end{abstract}


\maketitle



\section{Introduction} \label{sec:intro}

In the decades since the first exoplanet detections around a pulsar \citep{Wolszczan:1992aa} and a sun like star \citep{Mayor:1995aa}, there has been a crescendo in exoplanet research. Apart from the search for extra-terrestrial life, these discoveries have also changed our view of planet formation. Whereas previously the size distribution of planets around stars were thought to mirror that in our own solar system, the detection of \emph{51 Pegasi b} exposed loopholes in the theory of the evolution and formation of planets from protoplanetary disks. Campaigns such as Spitzer, Hubble, HARPS, JWST, and WFIRST have been launched with the goal of detecting exoplanets to: (1) address the possibility of extra-terrestrial life, (2) provide a window into the formation and evolution of stellar-planetary systems from gas/dust clouds and, (3) understand the interaction between planetary atmospheres and the parent star. Certainly, our view of planetary habitability is ``Earth-centric'' and hence key criteria are (a) the location in the circumstellar habitable zone (CHZ) of the parent star, (b) carbon-based life supporting chemistry, and the presence of liquid water and, (c) the existence of plate tectonics in order to modulate the carbonate-silicate cycle \citep[e.g.,][]{Korenaga:2010}. 

\citet{Seager:1998aa} discussed the effect of strong irradiation of an exoplanet atmosphere by stellar light and how it can be modeled to analyze the emergent spectra (i.e., the planet to star flux ratio) of the planet to study its atmosphere \citep{Charbonneau:1999aa, Cameron:1999aa}. \citet{Richardson:2007aa} studied the emergent spectrum of exoplanet HD209458b when it was in its secondary eclipse phase and discussed the possible presence of silicate clouds and the absence of water vapor in its atmosphere.

\citet{Seager:2000aa} demonstrated the utility of examining transmission spectra during the primary eclipse to study exoplanet atmospheric composition. The detection of the presence of sodium in the atmosphere of HD209458b by \citet{Charbonneau:2002aa}, paved the way for further detections and analyses of exoplanet atmospheres using transmission spectroscopy. For example, \citet{Vidal-Madjar:2003aa, Vidal-Madjar:2004aa} detected an extended atmosphere along with the presence of hydrogen, oxygen and carbon in the atmosphere of HD209458b, and \citet{Liang:2004aa} showed that due to extreme irradiation from stellar light, ``hot Jupiters'' have insignificant amounts of hydrocarbons relative to Jupiter and Saturn. \citet{Wit:2013aa} describe a model to constrain exoplanet mass using transmission spectroscopy, which traditionally requires radial velocity measurements, and demonstrated their approach for HD189733b.

The location of an exoplanet in the CHZ depends on the physical parameters of the system, and we have shown that these parameters can be determined directly from the data \emph{without} the canonical use of model fitting \citep{Sahil:EXO}. Probing the atmosphere of the detected exoplanet can reveal its chemical composition and hence characterize sufficient conditions for habitability. Due to their large size and orbital orientation, HD209458b \citep{Charbonneau:2000aa} and HD189733b \citep{Bouchy:2005aa} are two of the most studied exoplanets. Although ``hot Jupiters'' are amongst the most frequently detected exoplanets, a key detection target remains Earth-sized planets in the CHZ of a star. \citet{Ehrenreich:2006aa} have developed an atmospheric model relevant to such exoplanets and discussed their detectability based on, among other physical attributes, features from their transmission spectrum.

Due to their higher relative abundance and their strong infrared (IR) and near-IR absorption lines, H$_\textrm{2}$, CO, H$_\textrm{2}$O, CH$_\textrm{4}$, NH$_\textrm{3}$, along with atomic Na and K, are amongst the most studied chemical species in exoplanet atmospheres. However, even these seemingly simple atoms/molecules pose significant challenges in terms of modeling their relative contribution to the thermal and chemical profiles of an exoplanet atmosphere. A common assumption in most models is that of atmospheric chemical equilibrium, but given the proximity of ``hot Jupiters'' to their host stars, this assumption can often lead to questionable conclusions,  {and a disagreement between models and observations \citep{Steinrueck:2019, Baxter:2021}}. \citet{Fortney:2010aa} developed atmospheric models for hot Jupiters based on data from transmission spectra that assumed chemical equilibrium, but concluded that non-equilibrium chemistry is essential to properly explain the data. Another important characteristic that needs to modeled is the presence/absence of clouds and, if present, the particle size distribution of cloud condensates { \citep{Drummond:2018, Keles:2019, Steinrueck:2021}}. 
Given the infancy of our understanding of cloud chemistry and physics on Earth \citep{Pierrehumbertbook}, it is a serous challenge to model exoplanet clouds and their implications \citep[][and refs. therein]{Fletcher:2014aa, Burrows:2014aa}.  In addition to atmospheric scattering and absorption, the opacities of different chemical species at extreme temperatures and pressures, collision-induced effects, dynamical transport, the presence or absence of thermal inversion, amongst other processes, are all operative.  \citet{Madhusudhan:2012aa} take a step towards improving parameter estimation from observations and provide an analytic framework to interpret observables from reflected light, such as polarization parameters, geometric albedo and scattering phenomena, under the assumption of a semi-infinite homogenous atmosphere.

In this paper, we first briefly describe our model-free method of exoplanet detection \citep{Sahil:EXO} and compute the location of an exoplanet with respect to its parent star. Next, we use these results to examine exoplanetary atmospheres in a completely new light.  Namely, we analyze what we call the \emph{emission spectrum}, which is similar to the transmission spectrum except that the planet is in its secondary eclipse phase, during which we study the atmospheric limb. Thus, not only does the approach have implications for the study of exoplanet atmospheres, but it provides information to study the physical and dynamical characteristics of the exoplanet including, among others, atmospheric dynamics, tidal-locking behavior, and the dayside-nightside redistribution of energy.  The advantage of the approach resides in the lack of model assumptions in the detection and observational scheme.

\begin{figure}
    \centering
    \includegraphics[trim = 0 0 0 0, clip, width = 0.5\textwidth]{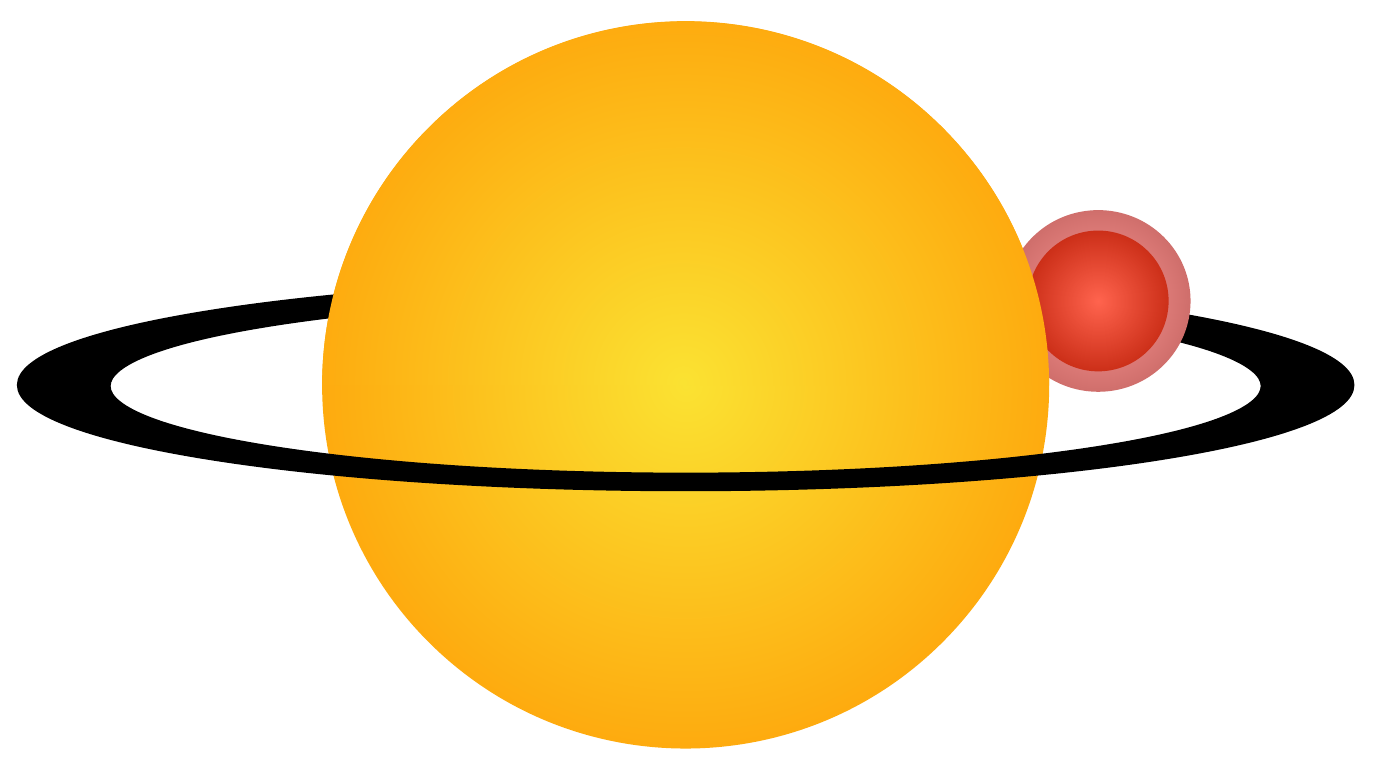}  	
    \caption{Sampling the planetary dayside limb (terminator) during a secondary eclipse event. While the widely studied emergent spectrum samples the area around the substellar point, sampling the limb allows one to study the upper atmosphere in the IR, where most spectroscopically active chemical species reside.}
    \label{fig:SecE}
\end{figure}

\section{Exoplanet Detection}\label{sec:detect}
Given a stellar series of spectral flux observations at regular time intervals, we construct a time-series for each wavelength in the spectrum.  For example, if the spectrum spans $L$ wavelengths, and we have $N$ observations, we construct $L$ time series each of length $N$. There are no {\em a-priori} assumptions regarding the temporal structure of these time-series, which are analyzed using a temporal multi-fractal approach. The multi-fractal scheme is ideally suited to this situation since the temporal fluctuations in flux at each wavelength can arise from photometric effects due to transit (both primary and secondary), atmospheric/telluric effects, instrumental noise, Doppler shifts, among other effects. Each of these phenomena have a characteristic timescale associated with them, which can be extracted by the multifractal approach. Finally, we determine the orbital timescales  $\tau_{12}$ (ingress/egress), $\tau_{23}$ (complete occultation) and $\tau_{14}$ (total transit), which correspond to the fluctuations in the emission spectra of the planet during secondary eclipse.   

We analyze all $L$ time series using Multi-fractal Temporally Weighted Detrended Fluctuation Analysis (MF-TWDFA) \cite[See][and references therein]{Sahil:MF, Sahil:EXO}, which does not \textit{a priori} assume anything about the temporal structure of the data. The approach has four stages, which we describe in Appendix \ref{Sec:method}.

\subsection{Physical Parameters for the Stellar-Planetary System}
The methodology described in Appendix \ref{Sec:method} has been shown to detect transiting exoplanets using NASA Spitzer mission data \citep{Sahil:EXO}.
Temporal multifractality is exhibited in the multiple time scales obtained from the data, which correspond to either the transit timescales ($\tau_{12}$, $\tau_{23}$ and $\tau_{14}$) or the wobble of the sensor on board the satellite ($\approx$ $55$ minutes). 

\begin{figure}
    \includegraphics[trim = 0 0 0 0, clip, width = 0.5\textwidth]{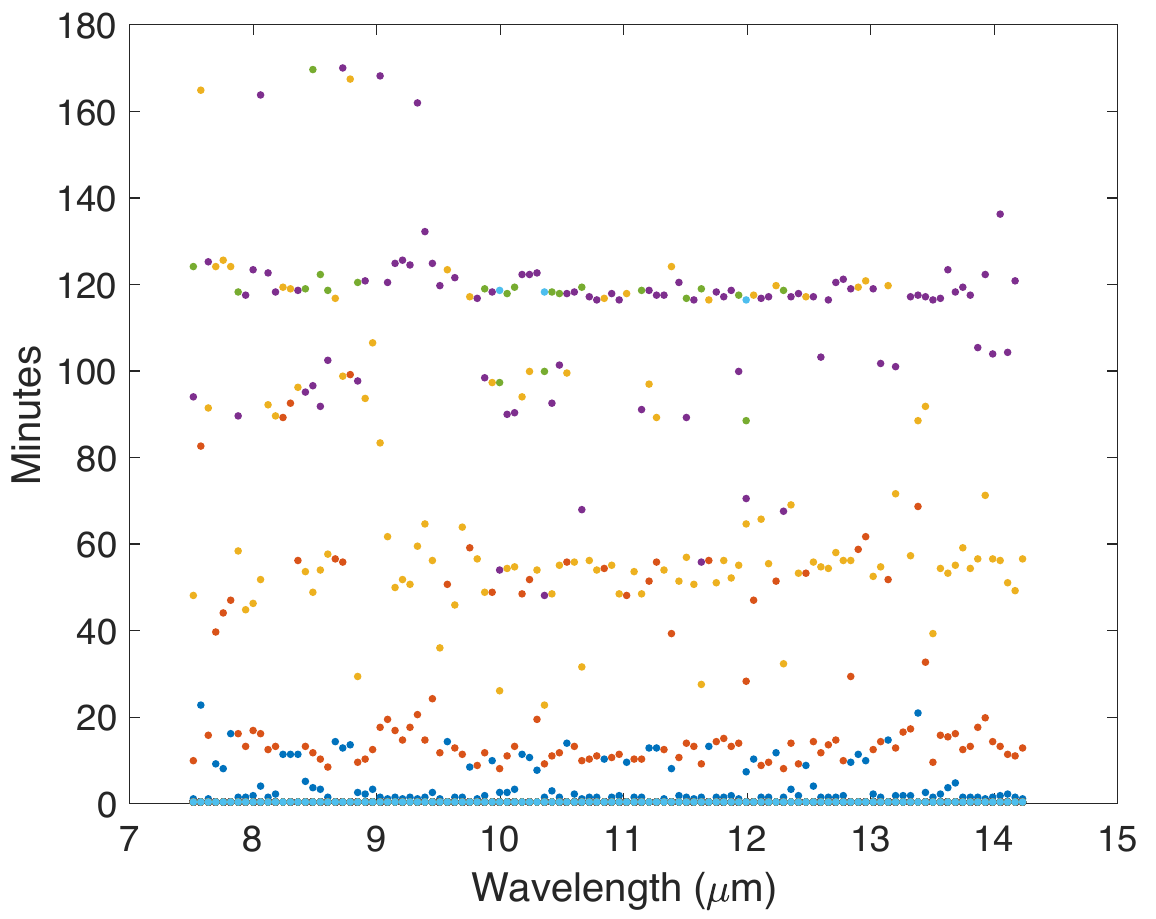}	  
    \caption{Reproduced from Figure 5 of \cite{Sahil:EXO}. The Spitzer based crossover times plotted for all wavelengths, for Night 2 (AOR-20645376, see Table 2 of \cite{Sahil:EXO}) for HD 189733b. All four significant timescales are robustly extracted using our method. (1) $\tau_{12} = $ 15.4804 $\pm$ 3.7660 minutes, (2) 55.0966 $\pm$ 5.8851 minutes, (3) $\tau_{23} = $ 87.0947 $\pm$ 2.7888 minutes, and (4) $\tau_{14} = $ 118.9671 $\pm$ 5.5764 minutes, where the uncertainty is one standard deviation about the mean. The timescales $\tau_{12}$ (ingress/egress), $\tau_{23}$ (complete occultation) and $\tau_{14}$ (total transit) correspond to the fluctuations in the emission spectra of the planet during secondary eclipse.}
    \label{fig:SpitzerD1}
\end{figure}

We showed that by combining the transit timescales with simple geometry, we can obtain the physical parameters of the system, such as the ratio of the planet to star radii 
 ($R_p / R_s$). Furthermore, this allows us to compute the decrease in intensity that one would observe if the planet were in a primary transit. From a sufficiently long set of observations, we can compute the orbital period of the exoplanet. This can then be used to compute the ratio of the orbital semi-major axis to the radius of the star or planet, density of the host star, planet surface gravity, among others. These physical parameters can then be used to examine if the exoplanet lies in the CHZ of its parent star. We note that these parameters are generally computed using the primary eclipse data, but we have demonstrated that one can utilize the secondary eclipse as well.

\section{Planetary Dayside Atmospheric Limb Retrieval}\label{sec:atmos}
Most exoplanet atmospheric studies examine either the dayside emergent spectrum when the planet is in a secondary transit or the transmission spectrum when the stellar flux passes through the atmosphere of the planet, sampling mainly its nightside limb.

While planets are often thought of as objects with sharp, well-defined, boundaries, the atmosphere blurs this boundary to varying degrees. During a primary transit, the stellar light passes through the atmosphere and, depending on its chemical composition, a wavelength dependent variation in the size of the planet is observed. This is due to the fact that the exoplanet's atmosphere, comprised of different atomic/molecular species becomes opaque to stellar light at different wavelengths, which correspond to atomic/molecular transitions. Therefore, this wavelength-dependent variation in the size of the exoplanet can be used to study exoplanet atmospheric properties \citep{Brown:2001aa}. Figure \ref{fig:SpitzerD1} shows the variation in transit timescales observed for HD189733b, which is equivalent to the variation in the size of the exoplanet.

\begin{figure}
    \includegraphics[trim = 40 10 10 10, clip, width = 0.5\textwidth]{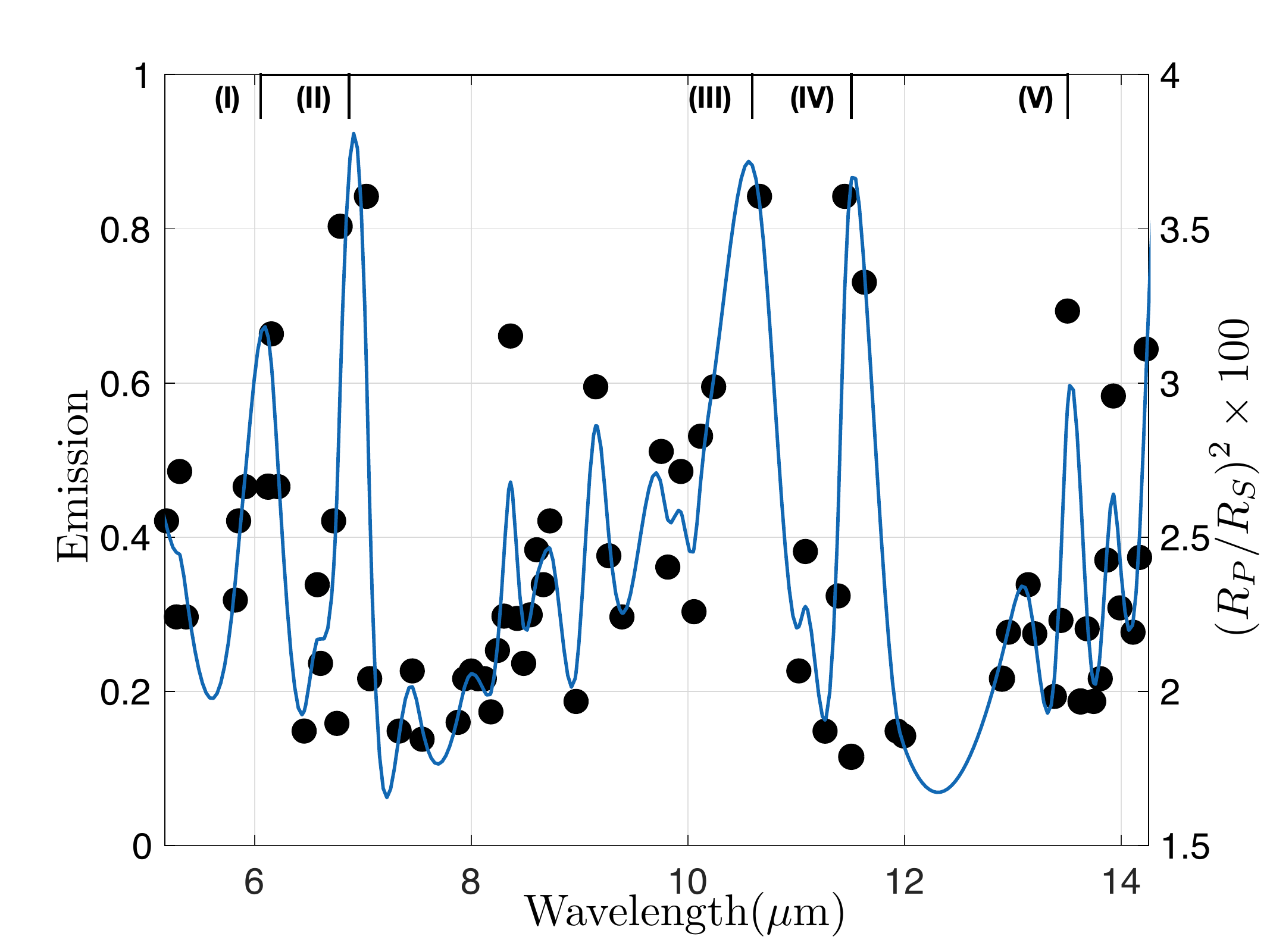}
    \caption{An emission spectrum (black dots) for the exoplanet HD189733b, showing the emission due to chemical species present in the exoplanet's atmosphere. This spectrum is sampling the planetary limb (terminator) during its secondary transit. We detect 5 peaks in the spectra, corresponding to (\RN{1}) the presence of water vapor, (\RN{2}) and (\RN{3}) ammonia, (\RN{4}) evidently a shifted ammonia peak and (\RN{5}) the presence of carbon dioxide. The ammonia peaks at $\sim6.9\mu$m, $10.5\mu$m and $11.6\mu$m are significant at $2.5\sigma$ level, the water vapor peak at $6.2\mu$m and the carbon dioxide peak at $13.5\mu$m are significant at $1.7\sigma$ level. The scale on the right axis denotes the apparent area of the planet with respect to its parent star. The blue curve is a cubic spline.}
    \label{fig:RadiiRatio}
\end{figure}

The size variation ($R_p^2/R_s^2$) of HD189733b is shown in Figure \ref{fig:RadiiRatio} with respect to its host star as a function of wavelength.  We note that the planet is in its secondary eclipse phase, namely moving behind the star. Generally, when examining the atmosphere during the secondary eclipse, one is interested in looking at the emergent flux from the planet,  which can be calculated by removing the stellar flux during a secondary eclipse from the out of eclipse flux. But this dayside emergent spectrum is only sampling the brightest parts of the planet or near the substellar point. In our case, the examination of the planetary limb (terminator) allows us to study the atmosphere of the exoplanet during its secondary eclipse and thus we are able to sample the atmosphere away from the substellar point. A comparison between the planetary limb atmospheric properties for the primary versus the secondary eclipse also allows one to study the flow structure and patterns on the exoplanet in great detail. Although theory predicts that for planets with an orbital period of less than 10 days tidal locking must be seen \citep{Winn:2010aa}, our method would also provide a means to study the tidal evolution of exoplanets.

As the planet moves behind the star, we observe the emitted spectral flux from the exoplanet (for hot-Jupiters) \citep{Burrows:2010aa}. Therefore, the peaks in the emission spectra correspond to the chemical species present in the atmosphere. While for the transmission spectra studies, the refraction of stellar light through a planet's nightside limb has to be accounted for, in the dayside limb's emission spectra this need not be an issue. Similarly, in transmission spectra one needs to decide on the cloud structure in the planet's atmosphere \citep{Seager:2000aa}, this is not required in our case. 

\subsection{HD189733b}
\citet{Deming:2006aa} first detected strong IR/thermal emission from HD189733b at $16\mu$ wavelength during its secondary eclipse phase. They examined the shape of the secondary eclipse, but were not able to characterize the thermal structure of the atmosphere using the available data. Based on  simulations of the infrared transmission spectra, \citet{Tinetti:2007ab} predicted that water vapor and carbon monoxide are the dominant species absorbing in the mid-IR, given that thermal structure plays a minor role relative to the molecular abundances for transmission spectra. Based on models and observations of transmission spectra, \citet{Tinetti:2007aa}  showed the abundance of atmospheric water vapor, while also discussing how different model assumptions can lead to different conclusions.
\citet{Swain:2008aa} analyzed the transmission spectra to show the presence of CH$_\textrm{4}$ in the atmosphere, \citet{Kok:2013aa} detected CO on the dayside using high-resolution spectroscopy, {\citet{Zhang:2020} confirmed the detection of CO and H$_\textrm{2}$O,  \citet{Flowers:2019} used high-resolution spectroscopy correlated with three-dimensional GCMs to detect CO and H$_\textrm{2}$O, } and \citet{Grillmair:2008aa} examined the emergent spectrum to show the presence of atmospheric water and discussed the possibility of the observations revealing the atmosphere changing with time. They also showed that the models used to fit these observations required a low dayside-nightside redistribution of energy factor. { \citet{Damiano:2019} used Principal Component Analysis to confirm the presence of H$_\textrm{2}$O in the planet's atmosphere.} \citet{Pont:2013aa} detected sodium and potassium in the atmosphere, but also note that their interpretation of the transit data may not be unique given the various parameters used in, and assumptions associated with, the atmospheric models. Finally, \citet{Knutson:2007aa} discuss tidal locking and its effect on energy redistribution from dayside to nightside.  

Typically, models for exoplanet atmospheres are used to provide a framework to describe the observations  \citep[][and refs. therein]{Burrows:2014ab}. Using a large number of model parameters with very few observations has often led to poor constraints on spectral retrieval of exoplanet atmospheric compositions \citep{Fletcher:2014aa}. Given high inter-model variability due to the range of assumptions and modeling techniques, there is rarely a unique model for a set of observations \citep{Madhusudhan:2009aa, Winn:2010aa, Burrows:2014ab}. However, given a set of exoplanet observations, our method determines the intrinsic timescales exhibited by the data (Fig. \ref{fig:SpitzerD1}), which can then be converted to a wavelength-dependent size variation of the exoplanet with respect to its host star (Fig. \ref{fig:RadiiRatio}). Because each wavelength has a corresponding transit timescale, this provides a much more finely resolved spectrum in wavelength space. Several peaks in the spectrum stand out, the most prominent being: 1) a peak at $\sim6.2\mu$m, corresponding the abundance of water vapor on the dayside of the planet \citep{Grillmair:2008aa}, 2) three peaks at $\sim6.9\mu$m, $10.5\mu$m and $11.6\mu$m, corresponding to the presence of ammonia and, 3) a peak at $13.5\mu$m, corresponding to carbon dioxide in the atmosphere (spectral line references from the NIST database). Using a one-dimensional photochemical model \citet{Moses:2011aa} showed that, due to the highly irradiated atmosphere of HD189733b, ammonia would be present at mole fractions higher than what would be allowed by thermochemical equilibrium. { \citet{Barth:2021} showed the presence of NH$_\textrm{4}$$^+$ ions in the planets atmosphere due to ionization by stellar particles}. Another prominent feature in the spectrum is the transit depth for ammonia at $\sim 4\%${, consistent with the upper limit proposed by \citet{Kilpatrick:2020}.} A possible explanation for such a large transit depth would be its presence in the extended atmosphere corresponding to ongoing evaporation \citep{Lecavelier-des-Etangs:2010aa}. { Temporal variations in the transit depth have been studied before and have been proposed due to escaping atmosphere, or the spectral energy distribution or solar flares \citep{Lecavelier-des-Etangs:2012, Guo:2016, Chadney:2017}.} Since the dayside emergent spectra samples the region near the substellar point, it would explain the absence of this feature in such studies. We emphasize the magnitude of these features and that no model-fitting is required to ascertain the causal chemical species. 

\section{Conclusion}
A central motivation of exoplanetary science is to understand if we are alone in the universe. The first step is to find planets outside our own solar system, which was taken by \citet{Wolszczan:1992aa} and \citet{Mayor:1995aa}. These discoveries substantially impacted our views of planet formation and evolution. The second step towards finding extra-terrestrial life is to sort through the exoplanet database by probing their atmospheres for chemical species that may support life, such as the observation of sodium, hydrogen, oxygen and carbon on HD209458b \citep{Charbonneau:2002aa, Vidal-Madjar:2003aa, Vidal-Madjar:2004aa}. Nonetheless, observation of chemical species alone, such as oxygen, are only potentially sufficient conditions \citep[e.g.,][]{Wordsworth:2014aa}.

A central challenge in the observation of exoplanet atmospheres is the use of atmospheric model-fitting, which has lead to many contradictory conclusions \cite[e.g.,][]{Grillmair:2007aa, Tinetti:2007aa}. One problem is that atmospheric models to not have unique solutions for the temperature, pressure and concentration of chemical species. This is partly due to the different assumptions associated with each model, and partly due to the common requirement that the data be \emph{noise--free}. 

In this Paper, by harnessing a new approach to exoplanet detection that is independent of model-fitting, and hence is free from any assumptions associated with such models \citep{Sahil:EXO}, we have provided a framework for atmospheric studies.  The detection approach treats the data as a temporal multifractal in a manner that uses noise as a source of information, from which we obtain the key physical parameters of the star-planet system and hence the presence of a planet in the CHZ of its host star. 
Here, we extended that approach to study the exoplanet atmosphere from the same dataset. 
Firstly, we used the results from the detection phase to retrieve the size variation of the planet with respect to its star as a function of wavelength. Secondly, we analyze what we term the \emph{emission spectrum} of the planet, which is analogous to the transmission spectrum except that the planet is in its secondary eclipse phase. Importantly, this provides a new window to study the spectroscopic signatures of different chemical species observed in the planet's atmospheric limb, without having to worry about the refraction of light or cloud structure, as is the case with transmission spectra. By comparing results from other atmospheric studies such as the transmission and the emergent spectra of the planet, this can further reveal processes such as atmospheric flows, tidal-locking behavior and dayside-nightside energy redistribution. 

We have demonstrated this approach for exoplanet HD189733b, where, by using the wavelength dependent transit timescales of the planet in its secondary eclipse \citep[from][]{Sahil:EXO}, we construct the size variation of the planet with respect to its star as a function of wavelength. The exoplanet does not have a well-defined boundary, and the chemical species present in its atmosphere absorb some of the incident light from the star corresponding to their transitions, leading to a wavelength dependent size of the planet.  We showed that the abundance of water vapor, ammonia and carbon dioxide is exhibited by strong features in its spectrum (Fig. \ref{fig:RadiiRatio}). This result is free from any assumptions associated with different atmospheric models.  Moreover, the emergent spectra principally samples the region around the substellar point on the exoplanet, thereby constraining the data to the uppermost layers of the atmosphere. However, our method, because it is sampling the planetary limb, can probe the lower layers of the atmosphere during the secondary eclipse phase, as demonstrated by the emission spectrum in Figure \ref{fig:RadiiRatio}.

This approach provides a robust and unique framework to detect and study exoplanet atmospheres solely using data. By characterizing the stellar flux as a multifractal, and thereby using the noise as a source of information, we can study exoplanet atmospheric spectroscopic signatures. This method provides a systematic approach to constrain different atmospheric model assumptions thereby honing the understanding of observed composition.\\
\\
The support of NASA Grant NNH13ZDA001N-CRYO is acknowledged by both authors.  J.S.W. acknowledges Swedish Research Council grant no. 638-2013-9243.

\appendix

\section{Multi-fractal Temporally Weighted Detrended Fluctuation Analysis}\label{Sec:method}
We analyze all $L$ time series using Multi-fractal Temporally Weighted Detrended Fluctuation Analysis (MF-TWDFA) \cite[e.g.,][and references therein]{Sahil:MF, Sahil:EXO}, which does not \textit{a priori} assume anything about the temporal structure of the data.  The method closest to ours is Pooled Variance Diagrams \citep{PVD}, and in \cite{Sahil:EXO}, we discuss the key differences between that approach and MF-TWDFA, the stages of which are as follows.  

\begin{itemize}

\item[(1)] Construct a non-stationary {\em profile} $Y(i)$ of the original time series $X_i$ as, 
\begin{equation}
Y(i )\equiv \sum_{k=1}^{i} \left(X_k - \overline{X_k}~ \right), \qquad \text{where}\qquad  i = 1, ... , N.  
\label{eq:profile}
\end{equation}
The profile is the cumulative sum of the time series and $\overline{X_k}$ is the average of the time series.

\item[(2)] This non-stationary profile is divided into $N_s = \text{int}(N/s)$ non-overlapping segments of equal length $s$, where $s$ is an integer and varies in the interval $1<s\le N/2$.  To account for the possibility that $N/s$ may not be an integer, this procedure is repeated from the end of the profile and returning to the beginning, thereby creating $2 N_s$ segments.  

\item[(3)]  A point by point approximation $\hat{y}_{\nu}(i)$ of the profile is made using a moving window, smaller than $s$ and weighted by separation between the points $j$ to the point $i$ in the time series such that $\vert i - j \vert \le s$.
A larger (or smaller) weight $w_{ij}$ is given to $\hat{y}_{\nu}(i)$ according to whether $\vert i - j \vert $ is small (large) \citep[][]{Sahil:MF}. This approximated profile is then used to compute the variance spanning up ($\nu = 1,...,N_s$) and down ($\nu = N_s + 1,...,2 N_s$) the profile as
\begin{eqnarray}
\text{Var}(\nu, s) & \equiv & \frac{1}{s}  \sum_{i=1}^{s} \{ Y([\nu - 1]s + i)  \nonumber \\
&&- {\hat{y}}([\nu-1]s +i) \}^2 \nonumber \\
&&\text{for $\nu = 1,...,N_s$ ~~~~and} \nonumber \\
\text{Var}(\nu, s) & \equiv & \frac{1}{s}  \sum_{i=1}^{s} \{ Y(N-[\nu - N_s]s + i) \nonumber \\
&&- {\hat{y}}(N-[\nu-N_s]s +i)\}^2\nonumber \\
&&\text{for $\nu =  N_s + 1,...,2 N_s$.}
\label{eq:varTW}
\end{eqnarray}

\item[(4)]  Finally, a {\em generalized fluctuation function} is obtained and written as
\begin{equation}
F_q (s) \equiv \left[ \frac{1}{2 N_s} \sum_{\nu=1}^{2 N_s} \{ \text{Var}(\nu, s)\}^{q/2} \right]^{1/q}.
\label{eq:fluct}
\end{equation}
\end{itemize}

As we vary the time segment $s$, the behavior of $F_q (s)$ will vary for a given order $q$ of the moment taken, which is characterized by a generalized Hurst exponent $h(q)$ as 
\begin{equation}
F_q (s) \propto s^{h(q)} .  
\label{eq:power}
\end{equation}
When $h(q)$ is independent of $q$ the time series is said to be monofractal, in which case $h(q)$ is  equivalent to the classical Hurst exponent $H$. For $q$ = 2, regular MF-DFA and DFA are equivalent \citep[e.g.,][]{Kantelhardt:2002}, and  $h(2)$ can also be related to the decay of the power spectrum $S(f)$. If $S(f) \propto f^{- \beta}$, with frequency $f$ then $h(2) = (1 + \beta)/2$ \citep[e.g.,][]{Ding}.  For white noise $\beta$ = 0 and hence $h(2) = 1/2$, whereas for Brownian or red noise $\beta = 2$ and hence $h(2) = 3/2$. The dominant timescales in the data set are the points where the fluctuation function $\ell{\text{og}}_{10}F_2(s)$ changes slope with respect to $\ell{\text{og}}_{10}s$. At each wavelength a crossover in the slope of a fluctuation function is calculated if the change in slope of the curve exceeds a set threshold, $C_{th}$. Because the window length is constrained as $1 < s \le N/2$ \citep{Zhou:2010}, this approach is limited to time scales of $t \leq t_\textrm{up} =  N \Delta t/2$.

\bibliography{ExoAtmosRefs}

\end{document}